\documentclass[prd, aps, superscriptaddress, preprintnumbers, twocolumn, floatfix, nofootinbib]{revtex4}
\pdfoutput=1

\usepackage{amsfonts}
\usepackage{amsmath}
\usepackage{amssymb}
\usepackage{bm}
\usepackage{dcolumn}
\usepackage{graphicx}   
\usepackage[latin1]{inputenc}
\usepackage{latexsym}
\usepackage{rotating}
\usepackage{hyperref}
\usepackage{graphicx}
\usepackage{color}

\newcommand\be{\begin{equation}}
\newcommand\ba{\begin{eqnarray}}
\newcommand\ee{\end{equation}}
\newcommand\ea{\end{eqnarray}}

\begin{document}

\title{Trans-Planckian Censorship Conjecture and Early Universe Cosmology}

\author{Robert Brandenberger}
\email{rhb@physics.mcgill.ca}
\affiliation{Department of Physics, McGill University, Montr\'{e}al, QC, H3A 2T8, Canada}

\date{\today}

\begin{abstract}

 I review the {\it Trans-Planckian Censorship Conjecture} (TCC) and its implications for cosmology, in particular for the inflationary universe scenario. Whereas the inflationary scenario is tightly constrained by the TCC, alternative early universe scenarios are not restricted.

\end{abstract}

\maketitle

\section{Introduction} 
\label{sec:intro}

Most of our current scenarios of early universe cosmology (see e.g. \cite{RHBrev1}  for a comparative review of several scenarios) are based on effective field theories. However, as has recently been emphasized in the context of the ``Swampland'' program, only a small set of effective field theories are consistent with a theory which is complete in the ultraviolet, specifically superstring theory (see e.g. \cite{swamprevs} for reviews on the Swampland program). In particular, potentials $V(\varphi)$ for scalar fields $\varphi$ which dominate the energy density of the universe are constrained to obey the conditions \cite{Vafa2, Krishnan}
\ba \label{deSitter}
\frac{|V^{\prime}| m_{pl}}{V} \, &>& c_1 \,\,\,\, {\rm{or}} \nonumber \\
\frac{V^{\prime \prime} m_{pl}^2}{V} &<& - c_2 \, ,
\ea
where $c_1$ and $c_2$ are positive constants of order one, a prime denotes the derivative with respect to $\varphi$, and $m_{pl}$ is the four dimensional Planck mass. Thus, the potential either has to be sufficiently steep, or sufficiently tachyonic. These conditions rule out standard single field slow-roll inflation \cite{Stein}. Applied to late time cosmology they imply that dark energy cannot be a cosmological constant, and they lead to constraints on quintessence models \cite{Lavinia}.

In the context of early universe cosmology, the challenges for effective field theories are accentuated due to the expansion of space. In the standard effective field theory treatment of linear cosmological perturbations, the fluctuating fields are expanded in Fourier modes in comoving coordinates, and each mode is treated as an independent harmonic oscillator (see e.g \cite{MFB, RHBfluctsrev} for reviews of the theory of cosmological perturbations). If we now consider a mode which today has a wavelength in the range of current observations, then if we go back sufficiently far into the past, the wavelength of this mode can become smaller than the Planck length. If we follow the scale corresponding to the present day Hubble radius back into the past using Standard Big Bang cosmology until a temperature of $10^{16} {\rm{GeV}}$, the scale of particle physics Grand Unification and the energy scale of inflation in the simplest single field scenarios, then the physical wavelength is in the range of ${\rm{cm}}$, many orders of magnitude larger than the Planck scale. However, in the case of inflationary cosmology wavelengths increase exponentially during the period of inflation, and, as discussed in \cite{Jerome}, this leads to a serious problem for the applicability of the usual treatment of fluctuations if inflation lasts for a long period of time.  Inflationary cosmology provides a causal mechanism for the origin of structure according to which all fluctuation modes emerge at the initial time (e.g. the beginning of the period of inflation) in their local quantum vacuum state \cite{ChibMukh, Starob}. But for a long period of inflation the corresponding wavelengths are trans-Planckian even for modes which are probed today on cosmological scales. This problem was called the {\it Trans-Planckian challenge} for cosmological perturbations \footnote{I use the word ``challenge'' and not ``problem'' to indicate the a correct understanding of Planck-scale physics can lead to signatures which can be looked for in cosmological observations.}. It was shown, in fact, that, in the context of inflationary cosmology, the prediction of a roughly scale-invariant spectrum of fluctuations is not robust to changing the assumptions about the physics on trans-Planckian scales \cite{Jerome}. As mentioned in \cite{Starob2}, trans-Planckian effects can even prevent inflation from starting \footnote{There has been a lot of work on the Trans-Planckian issue for inflationary cosmology. For a review with references to the original works see e.g. \cite{JeromeRev}.}. Note that it is the expansion of the cosmological background which is the source of the Trans-Planckian problem. This is an effect which is not present in other applications of effective field theory techniques, and the arguments do not invalidate the usual applications of effective field theory (as discussed e.g. in \cite{Burgess}). 
 
The canonical variables describing cosmological fluctuations (fluctuations induced by matter inhomogeneities, called {\it scalar fluctuations} in the literature) and gravitational waves (tensor fluctuations) oscillate on sub-Hubble scales (wavelengths smaller than the Hubble radius $H^{-1}$, where $H$ is the expansion rate), and freeze out and grow in amplitude once the wavelength becomes larger than $H^{-1}$. Cosmological observations probe scales which started out sub-Hubble and propagated for a long period of time in the super-Hubble range. This is illustrated in Fig. 1 in the case of inflationary cosmology. A condition for the absence of a trans-Planckian problem for fluctuations in the observable range is that the corresponding wavelengths were always larger than the Planck length.

Motivated both by the swampland program and by the above considerations, Bedroya and Vafa recently put forwards the {\it Trans-Planckian Censorship Conjecture} (TCC) \cite{TCC}. which states that no effective field theory emerging from superstring theory can lead to a situation where fluctuation modes which were initially trans-Planckian ever exit the Hubble radius. This conjecture leads to \cite{TCC2} very strong constraints on possible inflationary models. On the other hand, alternative cosmologies such as bouncing and emergent models are consistent with the TCC, as will be discussed. 

In the following section I will review the TCC and provide a number of motivations for it \footnote{See \cite{Andriot} for a verification of the TCC in a supergravity context, and \cite{Suddho} for a derivation based on the swampland criteria, but see also \cite{Saito} for a more critical view.} . In Section III I then discuss the constraints on inflationary models which follow from the TCC. In Section IV I consider application to some alternative early universe models and to late time cosmolgy, and give a further discussion of the results.

In the following we will consider a homogeneous and isotropic background cosmology given by the space-time metric
\be
ds^2 \, = \, dt^2 - a(t)^2 d{\bf x}^2 \, ,
\ee
where $t$ is physical time and ${\bf x}$ are comoving spatial coordinates. The Hubble expansion rate is given by
\be
H(t) \, = \, \frac{{\dot{a}}}{a} \, .
\ee
We will work in natural units in which the speed of light, Planck's constant and the Boltzmann constant are set to 1.

\section{Trans-Planckian Censorship Conjecture (TCC)}

The TCC \cite{TCC} states that in all models consistent with superstring theory the situation can never arise that an initally trans-Planckian wavelength of a fluctuation mode grows to be super-Hubble. The mathematical statement of this condition is \cite{TCC}
\be \label{TCCcond}
\frac{a(t)}{a(t_i)} l_{pl} \, \leq \, H^{-1}(t) \,\,\, \forall t > t_i 
\ee
which must hold for any initial time. This condition states that the physical wavelength of a mode which at the initial time $t_i$ was equal to the Planck length cannot grow to ever become larger than the Hubble radius at some later time $t$. 

The TCC implies that all information about trans-Planckian scales remains hidden from the ``classical domain'', the domain where fluctuations grow and can classicalize, i.e. the super-Hubble region \footnote{Since in principle nonlinearities could also form on sub-Hubble scales and become effectively classical, the above form of the TCC can be viewed as a neccessary but not a sufficient condition to have physics safe from the trans-Planckian region.}.

In non-accelerating cosmological backgrounds the physical size of the Hubble radius grows more rapidly than the physical wavelength of any mode, and hence (\ref{TCCcond}) does not lead to any restrictions. Similarly, in bouncing cosmological models and emergent cosmologies without a phase of acceleration after the bounce (or after the emergent phase), the TCC is automatically satisfied as long as the energy scale $\eta$ of the bounce (or the emergent phase) is smaller than the Planck scale, since in this case the modes which exit the Hubble radius during the phase of contraction (in the case of a bouncing cosmology) or at the end of the emergent phase (in the case of an emergent cosmology) always had a physical wavelength larger than $l_{\rm{min}}$, where
\be
l_{\rm{min}} \, \sim \, m_{pl}^{-1} \bigl( \frac{m_{pl}}{\eta} \bigr)^2 \, .
\ee
For the inflationary scenario, however, tight constraints result \cite{TCC2} which will be discussed in the following section. Before moving on, however, I will provide some heuristic arguments to support a milder form of the TCC, namely the statement that the condition (\ref{TCCcond}) must be satisfied in any cosmological model in which an effective field theory description applies. In a model in which (\ref{TCCcond}) is violated, a valid analysis must include effects which go beyond a local effective field theory. In this form, the condition should hold for any approach to quantum gravity, not just for superstring theory.

One justification of the TCC is in analogy with Penrose's Cosmic Censorship (CCH) hypothesis \cite{Penrose} which applies to black holes, and according to which any time-like singularity must be hidden from the outside observer by an event horizon. General Relativity as an effective field theory admits solutions with charge greater than the mass for which the CCH would be violated. For such solutions, the outside observer would not be shielded from the singularity, and the Cauchy problem for time evolution would not be well defined. The CCH states that in any theory which is complete in the ultraviolet such pathological solutions of the effective field theory cannot arise. The TCC can be viewed as an extension of this postulate to momentum space, and replacing the singularity by the set of trans-Planckian wavelengths and the event horizon by the Hubble radius (the Hubble horizon). In the same way that in the case of black holes the external observer must be shielded from the singularity by the event horizon, in cosmology the observer having access to infrared modes (sub-Hubble modes) must be shielded by the Hubble horizon from the trans-Planckian modes (see \cite{talk} where this argument was presented).

A second justification is based on the non-unitarity of effective field theory in an expanding universe \cite{Weiss}. In an effective field theory, an ultraviolet cutoff on the physical momentum of the field modes is required which cannot be higher than the Planck scale. Since the physical momentum of modes redshifts, one is forced to add Fourier modes to the Hilbert space of the effective field theory as the universe expands, in order to maintain the same ultraviolet cutoff in physical coordinates. The TCC is required to shield the observed from the non-unitarity of the effective field theory \footnote{Once again, this is a necessary but not a sufficient condition for the non-unitarity to effect cosmological observables.}.

\begin{figure}[htbp]
\centering
\includegraphics[scale=0.53]{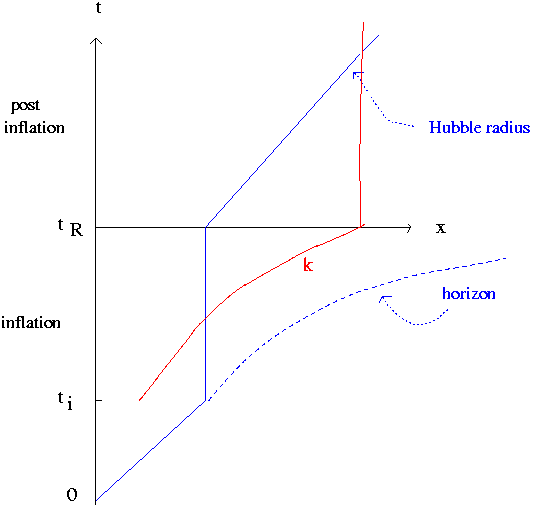}
\caption{Space-time sketch of inflationary cosmology. The vertical axis is time, the horizontal axis corresponds to physical distance. The inflationary phase of accelerated expansion lasts from time $t_i$ until time $t_R$. During this time interval and in the case of exponential expansion, the Hubble radius (the solid blue curve) is constant while the physical wavelength of a fixed comoving fluctuation mode (solid red curve marked by $k$) grows exponentially. Hence, modes can exit the Hubble radius. After inflation the universe evolves according to Big Bang cosmology, the Hubble radius grows linearly in time, and scales re-enter the Hubble radius.}
\end{figure}

Let us give a preview of the implications of the TCC for early universe cosmology. Figure 1 represents a space-time sketch of inflationary cosmology \cite{Guth}, the vertical axis being time and the horizonal axis representing physical distance. The phase of inflation (taken to be exponential in this figure) lasts from time $t_i$ until time $t_R$. After $t_R$ the universe evolves as in Standard Big Bang cosmology. The solid blue curve indicates the Hubble radius which is constant during the inflationary phase (note that the causal horizon grows exponentially during this phase), but the wavelength of a fluctuation mode (indicated by the solid red curve marked $k$ grows exponentially. Thus, modes which initially wave a very small wavelength are stretched and eventually re-enter the Hubble radius at late times as fluctuation modes which can be probed in cosmological observations. It is clear from this figure that if the period of inflation lasts a long time, there is the danger that modes which we observe today originate with a wavelength smaller than the Planck length. Thus, the TCC will have a major impact on inflationary cosmology.

As will be discussed in Section IV, some bouncing cosmologies (see e.g. \cite{bouncerev} for a recent review on bouncing cosmologies) can provide an alternative to cosmological iinflation for solving the horizon and flatness problem of Standard cosmology, and for providing a causal mechanism to generate cosmological fluctuations with a roughly scale-invariant spectrum.  Figure 2 depicts a space-time sketch of the resulting cosmology. The vertical axis represents conformal time $\tau$, and in this case the horizontal axis is comoving distance. The bounce time is denoted by $\tau_B$. For $\tau < \tau_B$ the universe is contracting, for $\tau > \tau_B$ space is expanding. To obtain a bouncing cosmology there has to be new physics which operates near the bounce point (between times $\tau_{B-}$ and $\tau_{B+}$. The comoving Hubble radius $|{\cal{H}}|^{-1}$ decreases in the contracting phase, and increases after the bounce. In comoving coordinate, the wavelength $\lambda$ of a fluctuation mode is fixed. Modes corresponding to fluctuations probed in cosmological observations originate at early times in the contrating phase with a wavelength similar to the current wavelength, and as long as the energy scale of the bounce is less than the Planck length, no mode which exits the Hubble radius ever and exit the Hubble radius. Hence, the TCC does not impose any constraints on an effective field theory description of fluctuations in a bouncing cosmology.
\begin{figure}[htbp]
\centering
\includegraphics[scale=0.4]{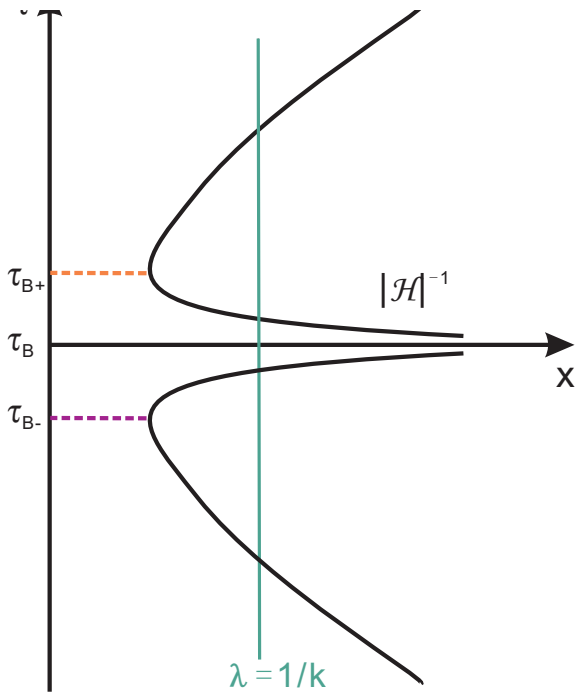}
\caption{Space-time sketch of a bouncing cosmology. The vertical axis is conformal time, the horizontal axis corresponds to comoving distance. The bounce happens at time $\tau_B$. Before that time, the universe is contracting, afterwards it is expanding. The comoving Hubble radius $|{\cal{H}}|^{-1}$ (shown as a solid black curve) decreases in the contracting phase, and thus scales (which have constant comoving wavelength) can exit the Hubble radius as they do in inflationary cosmology. However, the physical wavelength of the fluctuation mode is larger than the Planck length at all times, assuming that the energy scale of the bounce is smaller than the Planck scale.}
\end{figure}

As will also be discussed in Section IV, emergent universe scenarios can also provide an alternative to the theory of cosmological inflation. In these scenarios, the expanding phase of Big Bang cosmology emerges from an early phase which cannot be described by standard effective field theory. It might be a static Hagedorn phase of the gas of strings \cite{BV}, or a topological phase as recently suggested in \cite{Vafa3}. The space-time diagram in such a scenario is sketched in Figure 3. Here, the vertical axis is time and the horizontal axis corresponds to the physical distance. The time $t_R$ is when the transition from the initial phase to the radiation phase of Standard cosmology takes place. The solid blue curve represents the Hubble radius which starts out an infinity if the initial phase is static, and rapidly shrinks to a microscopic value as the transition time is approached. The red curves labelled $k_1$ and $k_2$ indicate the wavelengths of two fluctuation modes which exit the Hubble radius at times $t_i(k_1)$ and $t_i(k_2)$, respectively. Like in the case of a bouncing cosmology, as long as the energy scale of the initial phase is lower than the Planck scale, no mode which is ever super-Hubble was trans-Planckian at any earlier time. Hence, the TCC imposes no constraints on the effective field theory description of fluctuations.
\begin{figure}[htbp]
\centering
\includegraphics[scale=0.45]{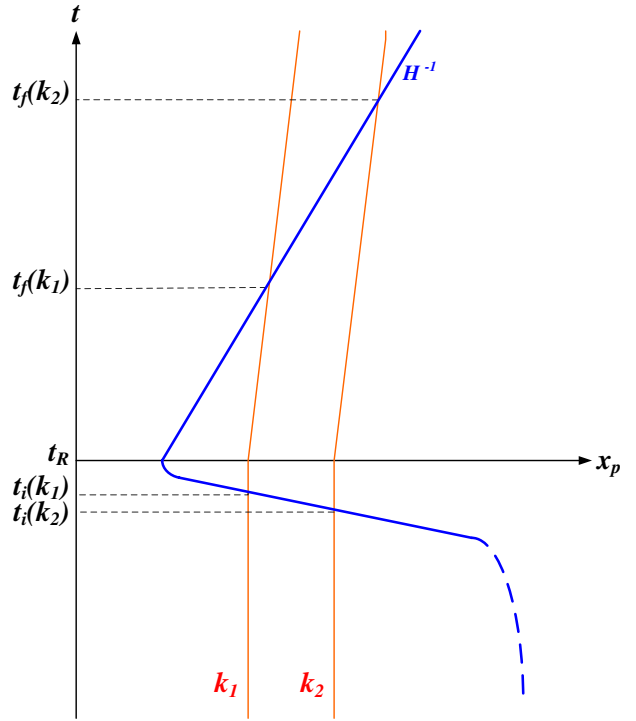}
\caption{Space-time sketch of an emergent cosmology. The vertical axis is time, the horizontal axis corresponds to physical distance. The transition between the early emergent phase (modelled here as being quasi-static) and the radiation phase of Standard Big Bang cosmology occurs at the time $t_R$. The Hubble radius is given by the solid bllue curve, while the two red curves labelled by $k_1$ and $k_2$ are the physical wavelengths of two comoving fluctuation modes. The physical wavelength of the fluctuation modes are larger than the Planck length at all times, assuming that the energy scale during the emergent phase (which sets the value of the Hubble radius at the time $t_R$) is smaller than the Planck scale.}
\end{figure}

\section{TCC and Inflationary Cosmology}

In this section we start by reviewing the constraints derived in \cite{TCC2} on standard slow roll inflation. If inflation is to be consistent with the TCC, then there is a conservative upper bound on the duration of the inflationary phase by demanding that the Planck length at the time $t_i$, the beginning of the inflationary phase, does not become larger than the Hubble length at the time $t_R$, the end of the period of accelerated expansion (after that time no scales will exit the Hubble radius), i.e.
\be \label{upper}
\frac{a(t_R)}{a(t_i)} l_{pl} \, < \, H^{-1}(t_R) \, .
\ee
Depending on the evolution of the universe before the onset of inflation, more stringent constraints can be derived, as was done in \cite{Edward} assuming that the inflationary period is preceded by a phase of radiation-dominated expansion.

On the other hand, if inflation is to provide a causal mechanism for the origin of structure in the universe, then fluctuations on the scale of the current Hubble radius $H^{_1}(t_0)$ must originate inside the Hubble radius at the beginning of inflation. This leads to a lower bound on the duration of inflation:
\be \label{lower}
\frac{a(t_i)}{a(t_0)} H^{-1}(t_0) \, < \, H^{-1}(t_i) \, ,
\ee
where $t_0$ denotes the present time.   

From Figure 1 it is easy to see that the consistency of the two bounds (\ref{upper}) and (\ref{lower}) will lead to an upper bound on the energy scale of inflation. The higher the energy scale of inflation, the smaller the Hubble radius is during inflation, and the less time (in units of Hubble expansion times) it takes for the Planck length to grow to become as large as the Hubble radius. On the other hand, it takes a longer time (again in units of the Hubble expansion time $H^{-1}$) for the waves whose wavelength equals the Hubble radius today to shrink to become to become sub-Hubble at the beginning of inflation. It is straightforward to work out that the constraint on the energy scale $\eta$ of inflation \footnote{In the case of inflation driven by the almost constant potential energy $V$ of a slowly rolling scalar field we have $\eta = V^{1/4}$.} (details are given in \cite{TCC2}) is
\be \label{bound}
\eta \, < \, \sqrt{3} m_{pl}^2 \bigl( T_{eq} T_0 \bigr)^{1/2} \, \sim \, 6 \times 10^8 {\rm{GeV}} \, ,
\ee
In the above, $T_{eq}$ and $T_0$ are the temperatures at the time of equal matter and radiation and the present time, respectively, and $m_{pl}$ is the Planck mass. This energy scale should be compared to the scale usually assumed in simple inflationary models in which the amplitude of the induced cosmological perturbations agrees with the observed value and is of the order $10^{15} {\rm{GeV}}$, the scale of particle physics Grand Unification. We thus see that inflationary models consistent with the TCC require a very low scale of inflation. Note that this constraint is independent of how inflation is obtained, and it also applies to {\it warm inflation} models \cite{warm}, models which can be made to be consistent with the swampland constraints \cite{warm2}, and to other inflationary models which can be made to be consistent with the swampland constraints (see e.g. \cite{Vahid2} for a discussion of some of these models). In particular, the TCC also applies to models in which inflation is obtained without making use of scalar matter fields.

Since the amplitude of the spectrum of primordial gravitational waves generated during inflation is set by the energy scale of inflation, or equivalently by the value of the Hubble expansion rate $H$ during inflation, the constraint (\ref{bound}) implies that the predicted amplitude \cite{Starob} of the dimensionless power spectrum ${\cal{P}}_h(k)$ of primordial gravitational waves is negligible \footnote{For an even tighter bound, derived in a less general context, see \cite{Kadota}.}:
\be
{\cal{P}}_h(k) \, \sim \, \bigl( \frac{H(k)}{m_{pl}} \bigr)^2 \, \sim \, 10^{-40} \, ,
\ee
where $H(k)$ is the value of $H$ when the mode $k$ crosses the Hubble radius (which is nearly scale-independent for nearly exponential expansion during inflation). Note that given this bound on the amplitude of the primordial gravitational waves, the spectrum of gravitational waves will be dominated by secondary effects.

Since in standard slow roll inflation models the dimensionless power spectrum ${\cal{P}}_{\zeta}$ of cosmological perturbations is given in terms of the {\it slow roll parameter} $\epsilon$ by
\be
{\cal{P}}_{\zeta} \, \sim \, \frac{1}{16 \epsilon} {\cal{P}}_h(k) \, ,
\ee
then, to obtain the observed amplitude of $10^{-9}$, a value of
\be
\epsilon \, < \, 10^{-32} 
\ee
is required. In the case of inflation driven by a scalar field, then to satisfy this condition requires extreme fine tuning of the potential, in particular if one wants a scenario in which the slow roll trajectory is a local attractor in initial condition space (see \cite{RHBrevIC} for a review on the initial condition issue in inflationary cosmology).

In the above, it was assumed that the period of inflation is almost exponential and that the transition to the radiation phase of Standard cosmology is instantaneous. Allowing for a non-standard post-inflation history or a multi-field inflation scenario leads to slightly weaker constraints \cite{Mukh}, but this benefit is at the cost of adding extra physics. Replacing almost exponential inflation by power law inflation also leads to relaxed constraints. As shown in \cite{Vahid}, the energy scale of inflation at the time when scales corresponding to presently measured cosmic microwave background (CMB) anisotropies exit the Hubble radius can be as high as $\eta \sim 10^{-5} m_{pl}$, which leads to a relaxed upper bound on the dimensionless amplitude of the gravitational wave power spectrum of ${\cal{P}}_h(k) < 10^{-20}$. 

As was already mentioned, stronger constraints on the scale of inflation can be derived if we consider specific pre-inflationary dynamics. In an expanding phase before inflation, the wavelength of fluctuations is increasing, and hence demanding that no scale which was trans-Planckian at the beginning of the pre-inflationary dynamics ever exited the Hubble radius leads to a stronger constraint than simply demanding that no scale which was trans-Planckian at the beginning of the period of inflation became larger than the Hubble radius \cite{Edward}. Specifically, assuming a radiation phase between the Planck density and the onset of inflation leads to the constraint
\be
\eta \, < \, 10^4 {\rm{GeV}} \, .
\ee

Note that in the context of inflationary cosmology, the TCC leads to other constraints on cosmology (see e.g. \cite{Wang}). Also, slightly weaker versions of the TCC have been studied \cite{weaker}.

\section{Discussion} \label{conclusion}

As has been reviewed here, the TCC is a condition for an effective field theory description of cosmology to be consistent. It leads to important constraints on any accelerating phase in the early universe, constraints which are independent of how this phase is obtained. The constraints force the energy scale of the hypothetical inflationary phase to be low (for concrete models in which this can be realized see \cite{Shafi}). Early universe models which do not involve an accelerating phase are hardly constrained by the TCC. However, the TCC also constrains models in which the period of canonical inflation is replaced by a period of k-inflation \cite{Will0} or a period of superluminal sound speed \cite{Will}. The TCC also applies to holographic cosmology \cite{Heliudson}, but it is not constraining in the context of nonlocal gravity \cite{Modesto}.

The Trans-Planckian Censorship Conjecture also has implications  for late time cosmology. In particular, it implies \cite{TCC} that Dark Energy cannot be a cosmological constant. In fact, any cosmology in which accelerated expansion continues arbitrarily far into the future is ruled out. As shown in \cite{TCC}, a metastable de Sitter state is consistent with the TCC provided that the lifetime is shorter than
\be \label{life}
T_{max} \, = \frac{1}{H} {\rm{log}} \frac{m_{pl}}{H} \, .
\ee 
In this sense, the TCC is a weaker condition than the {\it de Sitter} swampland condition (\ref{deSitter}) which rules out such cosmologies. Interestingly, the above time scale is the same at which the back-reaction of infrared cosmological perturbations indicates an instability of de Sitter space (see e.g. \cite{RHBrevBR} for a review), which in turn is the same as the {\it quantum break time} of de Sitter space in the approach of \cite{Dvali}.

Applied to a period of quintessential dark energy which lasts for a finite field range, the TCC leads to conditions on the slope of the scalar field potential which are \cite{TCC} similar but more specific than the criteria from the swampland conditions. In fact, in \cite{Bedroya}, the central role of the TCC in the swampland program was discussed.

Returning to the motivations for the TCC discussed iin Section II, we see that they are based on considering the evolution of fluctuations at the level of an effective field theory. Thus, a point of view that one could take is that the TCC does not necessarily imply that inflationary models with an energy scale which exceeds the bounds discussed in Section III are ruled out, but simply that they cannot be described at the level of an effective field theory. This is the view presented in \cite{Dvali2}. In this aspect there is a connection with the question of whether de Sitter solutions can be consistent with string theory. Based on the swampland criteria and on the TCC, the answer is ``no''. In fact, there are no-go theorems for the existence of de Sitter solutions at the level of an effective field theory (assuming that the internal manifold is time-independent) \cite{nogo} (There is, however, a lot of controversy on this issue. For an opposing point of view see e.g. \cite{false}.). However, going beyond an effective field theory approach it is possible to obtain de Sitter solutions, although they are unstable \cite{Dvali, Keshav}.

It is important to remember, however, that inflationary cosmology is not the only model which can explain the current date on CMB anisotropies and on the large-scale structure of the distribution of matter. As was discussed ten years before the development of inflationary cosmology \cite{SZ, PY}, given a roughly scale-invariant spectrum of almost adiabatic adiabatic fluctuations, the existence of the acoustic oscillations in the CMB angular power spectrum, and of baryon acoustic oscillations in the matter power spectrum follow. The inflationary scenario was the first scenario proposed which \cite{ChibMukh} yields such a spectrum based on causal physics, but it is not the only one. A bouncing cosmology in which the scales which are currenly observed exit the Hubble radius during a matter-dominated phase of contraction also yields such a spectrum, assuming that one starts (as one does in inflationary cosmology) with quantum vacuum perturbations \cite{Fabio}. This scenario, however, has anisotropy problems \cite{Peter} and is also in tension with the observed limit on the tensor to scalar ratio \cite{Jerome2}. The Ekpyrotic scenario \cite{Ekp} is a promising alternative to inflation since in this case the contracting phase is a global attractor in initial condition space. In the original Ekpyrotic scenario, it was necessary to add a spectator scalar field in order to obtain scale-invariant curvature fluctuations, but in the new version based on the addition of an S-brane to the low energy effective action \cite{newEkp} scale-invariant spectra of both cosmological perturbations and gravitational waves emerge directly. It would be interesting to study possible embeddings of this scenario in string theory.

Emergent cosmologies with holographic scaling of thermal correlation functions also provide an alternative to inflation for explaining the large scale structure of the universe. In the context of {\it String Gas Cosmology} \cite{BV}, the analysis of fluctuations was done in \cite{NBV}, and the prediction of a blue tilt of the spectrum of gravitational waves emerged \cite{BNPV}. Since this scenario is based on string theory, the swampland and TCC constraints are trivially satisfied. However, a good understanding of the emergent phase is still missing, but see \cite{Vafa3, us} for some promising approaches.

To end with a brief outlook: the tensions between the swampland constraints (and in particular the TCC) and the inflationary scenario indicate that it may be more promising to look beyond the inflationary scenario to obtain a good theory for the very early universe.

\section*{Acknowledgement}

\noindent The research at McGill is supported in part by funds from NSERC and from the Canada Research Chair program. RB is grateful for hospitality of the Institute for Theoretical Physics and the Institute for Particle Physics and Astrophysics of the ETH Zurich.

\end{document}